\documentstyle[prl,aps,twocolumn,floats,epsf]{revtex}

\newcommand{\plb}[2]{{\em Phys. Lett.}              {\bf #1B}, #2 }
\newcommand{\npb}[2]{{\em Nucl. Phys.}              {\bf B#1}, #2 }

\newcommand{\prt}[2]{{\em Phys. Rev.}               {\bf D#1}, #2 }
\newcommand{\pru}[2]{{\em Phys. Rev. Lett.}         {\bf  #1}, #2 }
\newcommand{\zpc}[2]{{\em Z. Phys.}                 {\bf C#1}, #2 }

\newcommand{\epc}[2]{{\em Eur. Phys. J.}            {\bf C#1}, #2 }

\newcommand{\etal}{{\em et al.}}

\newcommand{\as}{\hat\alpha_s}

\newcommand{\be}{\begin{equation}}
\newcommand{\ee}{\end{equation}}
\newcommand{\ba}{\begin{eqnarray}}
\newcommand{\ea}{\end{eqnarray}}

\def\ms{\mbox{{\footnotesize{$\overline{\rm MS}$}}} }

\newcommand{\gsim}{\buildrel > \over {_\sim}}

\def\al2{\frac{\alpha^2}{\pi^2}}

\begin{document}

\preprint{
\noindent
\hfill
\begin{minipage}[t]{3in}
\begin{flushright}
\vspace*{2cm}
\end{flushright}
\end{minipage}
}

\draft

\title{Hadronic Loop Corrections to the Muon Anomalous Magnetic Moment}
\author{Jens Erler$^{a}$ and Mingxing Luo$^{a,b}$}
\address{$^{a}$Department of Physics and Astronomy, University of Pennsylvania,
Philadelphia, PA 19104-6396, USA \\
$^{b,}$Zhejiang Institute of Modern Physics, Department of Physics,
Zhejiang University, Hangzhou, Zhejiang 310027, P R China}

\date{December 2000}

\maketitle

\begin{abstract}
The dominant theoretical uncertainties in both, the anomalous magnetic moment 
of the muon and the value of the electromagnetic coupling at the $Z$ scale,
$M_Z$, arise from their hadronic contributions. 
Since these will ultimately dominate the experimental errors, we study 
the correlation between them, as well as with other fundamental parameters.
To this end we present analytical formulas for the QCD contribution
from higher energies and from heavy quarks. Including these correlations 
affects the Higgs boson mass extracted from precision data.
\end{abstract}


The magnetic moment of the electron, $g_e$, provides the best determination of 
the fine structure constant, but is currently not measured precise enough 
to give a sensitive probe of electroweak physics which is suppressed by 
a factor $(\alpha/\pi) m_e^2/M_W^2 \sim {\cal O}(10^{-13})$.  Meaningful (but 
still weak) bounds on new physics contributions can be set only if they enter 
without loop suppression.  On the other hand, electroweak contributions to 
the anomalous magnetic moment of the muon~\cite{Kinoshita90},
$a_\mu = (g_\mu - 2)/2$, are enhanced by a factor 
$m_\mu^2/m_e^2 \sim 4\times10^4$, which renders them sizable enough to be 
detectable at the ongoing E821 experiment at the AGS at BNL.
($a_\tau$ has not yet been observed experimentally). E821 already 
reduced the experimental error to $\pm 1.6 \times 10^{-9}$~\cite{Brown01}. 
The anticipated final error of about $\pm 0.4\times 10^{-9}$ will mean a factor
of 20 improvement relative to previous results~\cite{Groom00}. $a_\mu$ 
therefore provides a good laboratory to test the Standard Model (SM) and probe 
theories beyond it~\cite{Czarnecki00}. For example, scenarios of low-energy 
supersymmetry with large $\tan\beta$ and moderately light superparticle masses 
can give large contributions to $a_\mu$~\cite{Lopez94}.

Unfortunately, the interpretation of $a_\mu$ is compromised
by a large theoretical uncertainty introduced by hadronic effects. 
Two and three loop vacuum diagrams containing light quark loops cannot be
calculated reliably in perturbative QCD (PQCD). Instead they are obtained by
computing dispersion integrals over measured (at low 
energies) and theoretical (at higher energies) hadronic cross sections.
At two loop~\cite{Gourdin69},
\be
   a_\mu (\mbox{had; 2-loop}) = \left(\frac{\alpha\; m_\mu}{3\pi}\right)^2
   \int_{4m_\pi^2}^\infty \frac{ds}{s^2} \hat{K}\left(s\right) R(s),
\label{eq9}
\ee
where $R(s)$ is the cross section of $e^+e^- \to \mbox{ hadrons}$, normalized
to the tree level cross section of $e^+e^- \to \mu^+\mu^-$, and where
\be
\hat{K}(s) = \int_0^1 dx \frac{3 x^2(1-x)} {1-x + x^2 m_\mu^2/s}.
\label{ks}
\ee
The uncertainty introduced by this procedure is significantly larger than 
the anticipated experimental one. For example, Ref.~\cite{Davier98} quotes 
an error of $\pm 0.67 \times 10^{-9}$, while other 
evaluations~\cite{Jegerlehner95,Alemany98,Jegerlehner99} give larger 
uncertainties. An analogous uncertainty occurs in the QED coupling constant, 
$\hat\alpha(\mu)$, (\ms quantities will be marked by a caret), preventing 
its precise theoretical computation from the fine structure constant, $\alpha$,
for $\mu \gsim 2 m_{\pi^0}$. Knowledge of $\hat\alpha (M_Z)$ is indispensable 
for the extraction of the Higgs boson mass, $M_H$, from the mass of the $W$ 
boson, $M_W$, and the weak mixing angle, $\hat{s}^2_W$. Again one must rely on 
a dispersion integral to be taken over $R(s)$, but using a different kernel 
function. As a result, these hadronic uncertainties are strongly correlated 
with each other, and also with other fundamental SM parameters, such as 
the strong coupling constant and the heavy quark masses. 

To address these correlations we obtain compact analytical expressions for 
$a_\mu (\mbox{had})$ wherever possible, {\em i.e.\/} for the charm
and bottom quarks, as well as for the high energy integration region (above 
$1.8$~GeV) for light quarks where PQCD appears to be applicable. A similar idea
has been pursued in Ref.~\cite{Erler99}, for the calculation of $\hat\alpha (M_Z)$.

To lowest order in QED, the anomalous magnetic moments are lepton universal and
mass independent~\cite{Schwinger48},
\be
  a_\ell^{\rm 1-loop} = {\alpha\over 2\pi}.
\ee
To two-loop order one has,
\be
  a_\mu^{\rm 2-loop} = a_\mu^\mu + a_\mu^e + 
  a_\mu^\tau + a_\mu^b + a_\mu^c + a_\mu^{uds} + a_\mu^{\rm had},
\label{2loop}
\ee
where the lepton contributions~\cite{Sommerfield57} amount to 
$a_\mu^\mu + a_\mu^e + a_\mu^\tau = 4132.18 \times 10^{-9}$.
$a_\mu^b$ and $a_\mu^c$ are heavy quark contributions. 
To proceed, we used the expansion,
\be
   \hat{K}(s) = 1 + k_1(\sqrt s) \frac{m_\mu^2}{s} 
              + k_2(\sqrt s)\frac{m_\mu^4}{s^2} + {\cal O}({m_\mu^6\over s^3}),
\label{khat}
\ee
where
\be
k_1(x) = \frac{25}{4} + 3 \ln \frac{m_\mu^2}{x^2}, \hspace{10pt}
k_2(x) = \frac{291}{10} + 18 \ln \frac{m_\mu^2}{x^2}. 
\ee
In Eq.~(\ref{eq9}) the integration is over the imaginary part of the photon 
polarization function which is related to $R$ by 
$R(s) = 12\pi{\rm Im}\Pi(q^2+i\epsilon)$ and is taken along the real axis from
the quarkonium threshold to infinity. Analytic continuation allows us to 
integrate instead over the full function, $\Pi(s)$, along a circle of radius 
$s = \hat{m}_q^2 (\hat{m}_q^2)$ counterclockwise around the origin.  This 
avoids a complicated integration over resonances,
{\em i.e.\/} a region where perturbative QCD is not applicable,
at the expense of introducing a new uncertainty from the imperfect knowledge of
the \ms quark mass 
(see the discussion in the next paragraph).
With $\Pi(s)$ known up to 
${\cal O}(\alpha_s^2)$~\cite{Chetyrkin97}, we find for a quark of charge $Q_q$,
$$ a_\mu^q = {Q_q^2\over 4} \al2 \left\{ {\frac{m_\mu^2}{4 \hat{m}_q^2}} \left[
   {16\over 15} + {3104\over 1215} a_s + a_s^2 \times \right. \right. $$
\vspace{-7pt}
\be \left. \left. \left( 0.5099 + {2414\over 3645} n_l \right) \right] \right.
   \left. + {\frac{m_\mu^4}{16 \hat{m}_q^4}} \left[ {108\over 1225} - 0.1943\;
    a_s + \right. \right.
\label{amuq}
\ee
\vspace{-14pt}
$$ \left. \left. 3 \ln \frac{m_\mu^2}{\hat{m}_q^2} \left( {16\over 35} + 
   {15728\over 14175} a_s + 1.4123\; a_s^2 + {290179\over 637875} n_l a_s^2 
   \right) \right] \right\}, $$
where $a_s = \alpha_s(\hat{m}_q)/\pi$, $n_l = 3$ for charm, and $n_l = 4$ for 
bottom. Higher orders in $m_\mu^2/\hat{m}_q^2$ can safely be neglected. 
Terms of ${\cal O} (\alpha_s^2 m_\mu^4)$ can also be dropped unless they are 
logarithmically enhanced. 
Non-perturbative effects in 
the operator product expansion were also 
computed and found to be negligible.  
Using (here and in the following) 
$\alpha_s (M_Z) = 0.120$, $\hat{m}_c = 1.31$~GeV, and $\hat{m}_b = 4.24$~GeV, 
we find $a_\mu^c = 1.39 \times 10^{-9}$ and $a_\mu^b = 0.03 \times 10^{-9}$.

In contrast to the numerical integration over resonances, Eq.~(\ref{amuq}) is 
a simple and transparent representation of the heavy quark contribution. 
More importantly, the uncertainty implied by Eq.~(\ref{amuq}) is smaller.
To see this, notice first the excellent behavior of its $\alpha_s$ expansion
implying a small truncation error which we evaluated to be 
$\pm 0.05\times 10^{-9}$. The dominant uncertainty is induced via the \ms charm
mass, but it is {\em necessarily\/} smaller than in the conventional approach 
{\em regardless\/} of the available data in the resonance region: if one 
compares the two treatments --- integration over the real axis heavily relying 
on experimental results and integration over the circle contour relying only on
PQCD --- one can {\em determine\/} the heavy quark masses by demanding 
consistency.  This effectively amounts to deriving a {\em specific\/} QCD sum 
rule. The point we are making here, is that this is only one of a large number 
of possible sum rules, but not the one which uses the available information 
most efficiently. The charm mass extracted from the most efficient sum 
rule~\cite{Eidemuller01} (and other precise quark mass determinations)
can then be used for Eq.~(\ref{amuq}). An unweighted average of various
determinations yields~\cite{Erler99} $\Delta\hat{m}_c = \pm 0.07$~GeV and 
$\Delta a_\mu^c = \pm 0.16 \times 10^{-9}$. 
In the near future these uncertainties are likely to reduce 
to about $\pm 0.04$~GeV and $\pm 0.09 \times 10^{-9}$, respectively.
Similarly, $\Delta\hat{m}_b = \pm 0.11$~GeV, 
but the induced error in $a_\mu$ is negligible. 

The remaining terms in Eq.~(\ref{2loop}) are due to $u$, $d$, and $s$ quark
effects, which we separated into the contributions from $\sqrt{s}\geq \mu_0 = 
1.8$~GeV ($a_\mu^{uds}$) and $\sqrt{s} \leq \mu_0$ ($a_\mu^{\rm had}$). 
$a_\mu^{uds}$ can be written as an expansion in $m_\mu^2/\mu_0^2$. For the 
leading contribution we find,
\ba
   a_\mu^{uds;{\rm LO}} = {2\over 9} \al2 {m_\mu^2\over \mu_0^2}
      \left( 1 + B_1 + \sum\limits_{n=2}^\infty d_n B_n \right),
\label{amuudsci}
\ea
where $d_2 = 299/24 - 9 \zeta(3)$ and 
$d_3 = 58057/288 - 779 \zeta(3)/4 + 75 \zeta(5)/2$ are coefficients of the
Adler $D$-function, and
\ba
   B_n = \frac{1}{2\pi i} \oint_{|s|=\mu_0^2} \frac{ds}{s} \left( 
     1 - \frac{\mu_0^2}{s} \right) \left[ \frac{\alpha_s(-s)}{\pi} \right]^n,
\ea
which we compute with a 4-loop renormalization group improvement. 
For a representative value $\alpha_s (\mu_0)/\pi = 0.1$,
\ba
B_1 = 7.069 \times 10^{-2}, \hspace{20pt}
B_2 = 4.514 \times 10^{-3}, \nonumber \\
B_3 = 2.562 \times 10^{-4}, \hspace{20pt}
B_4 = 1.243 \times 10^{-5}. 
\ea
Notice, that $B_4$ is small enough, that even with the fourth order 
coefficient, $d_4$, unknown this treatment keeps the truncation error at
a negligible level. (For an estimate of the uncertainty of $d_4$, see 
Ref.~\cite{Erler99A}.) Denoting $k^\prime_1= k_1(\mu_0) - \frac{3}{2}$ and 
$k^\prime_2= k_2(\mu_0) - 6$, we find for the subleading contributions,
\ba
  {\pi^2\over \alpha^2} a_\mu^{uds;{\rm rem}} = 
     {2\over 9} a_s {m_\mu^2 \hat{m}_s^2(\mu_0^2)\over \mu_0^4} + a_s^2 
     {m_\mu^2\over \mu_0^2} G\left( {\mu_0^2\over \hat{m}_c^2(\hat{m}_c^2)} 
     \right) \nonumber \\
   + \left\{ (1 + a_s) {k^\prime_1\over 9} + a_s^2 \left[ \left(
     \frac{34}{27} - \zeta(3) \right) k^\prime_1 + \frac{3}{16} 
     \right] \right\} {\frac{m_{\mu}^4}{\mu_0^4}} \\
   + {2\over 3} \left\{ (1 + a_s) {k^\prime_2\over 9} + a_s^2 \left[ \left( 
     \frac{281}{216} - \zeta(3) \right) k^\prime_2 + {1\over 2}
     \right] \right\} {\frac{m_{\mu}^6}{\mu_0^6}}, \nonumber 
\ea
where in ${\cal O} (\as)$ we kept the small $s$ quark mass effect 
($\sim 3\times 10^{-12}$). $G(x)$ arises from virtual charm quark effects
inside a light quark loop (double bubble diagram). Even below threshold it can 
be well approximated as an expansion in $x$ despite the fact that 
$\hat{m}_c^2 < \mu_0^2$,
$$
  G(x) \approx {x\over 1215} \left[ \frac{3503}{75} - \frac{2\pi^2}{3} - 
    {88\over 5} \ln x + 2 \ln^2 x \right] $$
\be
  + {x^2\over 11340} \left[ \frac{1723}{420} - \ln x \right].
\ee
$G(x)$ also applies to $b$ quarks, but this contribution can be safely 
neglected. We find $a_\mu^{uds} = 4.38 \times 10^{-9}$ 
and the leading order resummation in Eq.~(\ref{amuudsci}) renders 
the truncation error negligible. 

Leading order non-perturbative contributions due to gluon and light quark 
condensates are given by, 
\be
   a_\mu^{uds;{\rm NP}} = {2\alpha^2\over 81} a_s^2 {m_\mu^2\over\mu_0^2}\left[
   {11\over 4}\frac{<a_s G G>}{\mu_0^4} - \frac{<m_s\bar{s} s>}{\mu_0^4}\right]
\ee
where the condensates are of order $\Lambda_{\rm QCD}^4$ and $- m_K^2 f_\pi^2$,
respectively. These terms are suppressed by two powers of $\alpha_s$ and they 
change $a_\mu$ by less than $10^{-12}$, an effect completely negligible. 
Effects from up and down quark condensates are suppressed by a further factor 
of $m_\pi^2/m_K^2$, and quartic mass terms are tiny, as well. 
Uncertainties from non-perturbative effects not accounted for by the operator 
product expansion, which are due to the transition from the data region to PQCD
at $\mu_0$, have been estimated in Ref.~\cite{Davier98} which quotes
$\pm 0.024\times 10^{-9}$.

We take the low energy contribution from Ref.~\cite{Davier98}, 
\be
  a_\mu^{\rm had} = (63.43 \pm 0.60) \times 10^{-9},
\label{amuhad}
\ee
which includes a QCD sum rule improvement, PQCD down to a relatively low 
$\mu_0 = 1.8$~GeV, 
as well as additional information from $\tau$-decays.
The quoted error is not 
uncontroversial~\cite{Czarnecki00,Jegerlehner99} and needs to be confirmed. 
Note, however, that inclusion of the $\tau$-decay data decreases 
the difference between the SM prediction and the current experimental 
result~\cite{Brown01}. Our reason to use Ref.~\cite{Davier98} is that it
quantifies the correlation (69\%) with the corresponding result on 
$\Delta\alpha_{\rm had} = (56.53 \pm 0.83) \times 10^{-4}$. We also account
for the almost perfect anti-correlation with higher order hadronic 
uncertainties, as will be discussed below. The total two-loop quark 
contribution is 
$a_\mu^b + a_\mu^c + a_\mu^{uds} + a_\mu^{\rm had} = 69.23 \times 10^{-9}$. 

This completes the discussion of the terms appearing in Eq.~(\ref{2loop}).
The leading contribution at ${\cal O} (\alpha^3)$ is from the light-by-light
diagram containing an electron loop~\cite{Laporta93},
$$ a_\mu^{\rm e;lbl} = {\alpha^3\over \pi^3} \left[ {59\over 3} \zeta(4) + 
  4 \zeta(2) \left( \ln {m_\mu\over m_e} - 5 \right) - 3 \zeta(3) + {2\over 3}
   \right] $$
where we displayed only the leading term in explicit form, but the terms
suppressed by one or two powers of $m_e/m_\mu$ are significant, as well.
The contribution due to $a_\mu^{\rm e;lbl} = 20.9479\; \alpha^3/\pi^3 
= 262.54 \times 10^{-9}$ is almost four times larger than the entire 
hadronic contribution at ${\cal O} (\alpha^2)$. The corresponding contribution 
involving a muon loop is $a_\mu^{\rm \mu;lbl} = 0.3710\; \alpha^3/\pi^3
= 4.65 \times 10^{-9}$, and the $\tau$ contributes~\cite{Laporta93},
\be
   a_\mu^{\tau;{\rm lbl}} = {\alpha^3\over \pi^3}
   {m_\mu^2\over m_\tau^2} \left[ {3\over 2} \zeta(3) - {19\over 16} \right]
   = 0.03 \times 10^{-9}.
\ee
An evaluation of the hadronic light-by-light contribution yields~\cite{Hayakawa98},
\be
   a_\mu^{\rm had;lbl} = ( - 0.792 \pm 0.154) \times 10^{-9},
\label{amulbl}
\ee
which is consistent with the finding in Ref.~\cite{Bijnens96}. 
Notice, hadrons and leptons contribute with opposite signs. 
Since these results
are based on model calculations, an independent confirmation would be desirable. 

The purely leptonic 3-loop vacuum polarization contribution is again dominated
by electron loops and given by $a_\mu^{\ell;{\rm vpol}} = 2.7294\; 
\alpha^3/\pi^3 = 34.21 \times 10^{-9}$. On the other hand, the contribution 
from two hadronic loops is suppressed by a factor $m_\mu^4/ 16 m_{\pi^\pm}^4$ 
and therefore small, $a_\mu^{\rm had;vpol} = (0.027 \pm 0.001) \times 
10^{-9}$~\cite{Alemany98}. Similar to our strategy at two loops, we separated 
the mixed leptonic-hadronic contribution into heavy quarks, light quarks 
($\sqrt{s}\geq \mu_0$), and light hadrons ($\sqrt{s}\leq \mu_0$). 
The $\tau$-hadronic contribution is of the order of $10^{-12}$ and negligible. 
As for the other leptons, we use the kernel functions from Ref.~\cite{Krause97}
and obtain for charm and bottom quarks,
$$ a_\mu^{\ell-q;{\rm vpol}} = \frac{\alpha^3}{\pi^3} Q_q^2  
   \frac{m_\mu^2}{4\hat{m}_q^2} \left[ \left( \frac{8}{5} + \frac{1552}{405} a_s
 \right) 
  \right.  $$
\vspace{-14pt}
\be
    \left. \left( \frac{110}{27} - \frac{\pi^2}{3}  
   +   \frac{23}{36} \ln \frac{m_\mu^2}{\hat{m}_q^2}
   + \frac{1}{9} \ln \frac{m_\mu^2}{m_e^2} \right) - \frac{1771}{675} \right] ,
\ee
which amounts to $-0.05 \times 10^{-9}$ and $-0.002\times 10^{-9}$, 
respectively. For the light quarks we obtain,
$$ a_\mu^{\ell-uds;{\rm vpol}} = \frac{\alpha^3}{\pi^3} {1 + a_s\over 9} \left[
   \frac{m_\mu^2}{\mu_0^2}\left( {371\over 9} - 4\pi^2 + {23\over 3} 
   \ln \frac{m_\mu^2}{\mu_0^2} + \right.\right.$$ 
\vspace{-14pt}
\be
  \left. {4\over 3} \ln \frac{m_\mu^2}{m_e^2} \right) + \frac{m_\mu^4}{\mu_0^4}
  \left( \frac{20359}{576}-{103\over 24}\pi^2 + 
  {509\over 72} \ln \frac{m_\mu^2}{\mu_0^2} + \right. 
\ee
\vspace{-14pt}
$$ \left. \left. {19\over 6} \ln \frac{m_\mu^2}{m_e^2} - 
   {5\over 24}\ln^2 \frac{m_\mu^2}{\mu_0^2}
   - 2  \ln \frac{m_\mu^2}{\mu_0^2} \ln \frac{m_\mu^2}{m_e^2} \right) \right]
   = - 0.15 \times 10^{-9}. $$

We found the vacuum polarization contribution arising from an electron in one 
loop and light hadrons in the other one, 
$a_\mu^{e-{\rm had;vpol}} = 0.97 \times 10^{-9}$, by constructing a simplified 
function $R(s)$ which reproduces the results in Ref.~\cite{Davier98}. 
$a_\mu^{e-{\rm had;vpol}}$ is almost completely ($ > 99.9\%$) correlated with 
$a_\mu^{\rm had}$ in Eq.~(\ref{amuhad}). On the other hand, 
$a_\mu^{\mu-{\rm had;vpol}} = -1.80 \times 10^{-9}$ is very strongly 
($\approx - 97\%$) anti-correlated with $a_\mu^{\rm had}$. The small 
uncorrelated error contributions are clearly negligible, and the correlated 
ones can be added (subtracted) linearly, slightly reducing the error in 
Eq.~(\ref{amuhad}) to $\pm 0.59\times 10^{-9}$. Including the uncertainty from 
the light-by-light contribution in Eq.~(\ref{amulbl}) we obtain 
$\pm 0.61\times 10^{-9}$ as the total hadronic error excluding parametric
uncertainties. 

Taking the other theoretical uncertainties mentioned earlier to
be 100\% correlated with the corresponding uncertainties in 
$\Delta\alpha_{\rm had}$ (which is a fit parameter),
we obtain a residual correlation of 55\% or $\pm 0.34\times 10^{-9}$
(the uncorrelated error is $\pm 0.51 \times 10^{-9}$).

The four-loop contribution~\cite{Kinoshita93}, 
$a_\mu^{\rm 4-loop} =$ $126.04 \alpha^4/\pi^4$ $= 3.67 \times 10^{-9}$, 
and the five-loop estimate~\cite{Karshenboim93}, 
$a_\mu^{\rm 5-loop} = 930   \alpha^5/\pi^5 = 0.06 \times 10^{-9}$,
are also included.

One-loop electroweak corrections due to $W$ and $Z$ boson loops are given 
by~\cite{Brodsky67},
\be
   a_\mu^{\rm EW;1-loop} = {G_F m_\mu^2\over 24\sqrt{2}\pi^2} 
                              (5 + (1 - 4 s^2_W)^2).
\label{1lew}
\ee
Two-loop corrections to Eq.~(\ref{1lew}) are significant due to large 
logarithmic contributions~\cite{Kukhto92,Czarnecki95,Czarnecki96},
\be
   {\alpha G_F m_\mu^2\over 24\sqrt{2}\pi^3}\left[ 41 
      - {124\over 3} s^2_W (1 - 2 s^2_W) \right] \ln {M_Z^2\over m_\mu^2}. 
\ee
The fermionic two-loop result, including some additional logarithms and 
sub-leading contributions, has been obtained in Ref.~\cite{Czarnecki95} for 
small and large values of $M_H$.  We constructed an interpolation formula for 
other values of $M_H$ which reads in units of 
$\alpha(m_\mu) G_F m_\mu^2/(3\sqrt{2}\pi^3)$,
\be 
   {13 + 6 \ln x \over 9} (1 - \omega) +
   x \left[ 3 + {\pi^2\over 3} + (\ln x + 1)^2\right] \omega, 
\label{interpol}
\ee
where $x = \hat{m}_t^2/M_H^2$ and $\omega = e^{ - 3.31849 x}$. 
Eq.~(\ref{interpol}) reproduces the exact result~\cite{Czarnecki95} for 
$M_H = \hat{m}_t$. 
The bosonic two-loop corrections have been obtained only for large values of 
$M_H$ and as an expansion in $s^2_W$~\cite{Czarnecki96}. We take the leading 
non-logarithmic contribution in the $M_H \rightarrow \infty$ limit ($a_{-2}$ in
Ref.~\cite{Czarnecki96}) as the uncertainty induced by sub-leading bosonic 
two-loop effects. The leading logarithms to three-loop order have also 
been computed~\cite{Degrassi98} and included in our analysis. We obtain 
$a_\mu^{\rm EW} = (1.52 \pm 0.03) \times 10^{-9}$. 

Summation of the various contributions gives
$(g_\mu - 2 - {\alpha\over\pi})/2 = 4506.28 \times 10^{-9}$.
Including our evaluation of $a_\mu$ into a global analysis of electroweak 
data yields,
\be
   {1\over 2}(g_\mu - 2 - {\alpha\over\pi}) 
   = (4506.35 \pm 0.37 \pm 0.51) \times 10^{-9},
\label{amuSM}
\ee
where the first error includes $\Delta\alpha_{\rm had}$ and all other 
parametric uncertainties, such as the $\pm 0.028$ error in
$\alpha_s (\mu_0)$ obtained from current global fits. The extracted Higgs mass 
from this fit is $M_H = 88_{-33}^{+49}$~GeV. The current experimental 
world average, $(4510.55 \pm 1.51) \times 10^{-9}$ (which differs by 
$2.6\sigma$ from Eq.~(\ref{amuSM})) needs to be included into the analysis, 
as well. This is achieved by combining the experimental error with the second error
in Eq.~(\ref{amuSM}) in quadrature (for a total of $1.59 \times 10^{-9}$).
The global fit then yields $4506.52 \pm 0.36$ (parametric) with a pull of
$2.5$. The crucial observation here is that now $M_H = 83_{-31}^{+47}$~GeV,
{\em i.e.\/} both, the central value and the uncertainty of $M_H$ decrease.
The increase in the precision in $M_H$ gained by properly correlating it to
$\Delta\alpha_{\rm had}$ (and assuming the SM) is almost identical to that 
provided by the $M_W$ measurement at the Tevatron Run~I.

To summarize, we carefully examined the error correlations between $a_\mu$ and 
other quantities entering electroweak tests, especially $\Delta \alpha (M_Z)$,
$\alpha_s$, and the quark masses. 
We derived new analytical results for the hadronic contributions and showed 
that the proper treatment discussed in this article has a significant effects 
on the extraction of $M_H$, which could (depending on the future experimental 
findings) become even more dramatic. 

{\bf Acknowledgements:}
It is a pleasure to thank P.~Langacker and W.~Marciano for valuable 
discussions. This work was supported in part by the US Department of Energy 
grant EY-76-02--3071, CNSF-10047005, and a fund for Trans-Century Talents.

\end{document}